\documentclass[aps,prx,twocolumn,a4,showkeys,10pt,longbibliography]{revtex4-1}

\usepackage{graphicx}
\usepackage[english]{babel}
\usepackage{amsmath}
\usepackage{float}
\usepackage{hyperref}
\usepackage{subfigure}
\usepackage{tikz}
\usepackage{physics}
\usepackage{breqn}
\usepackage{mathtools}
\usepackage{enumerate}
\usepackage{amssymb}

\newcommand*\circled[1]{\tikz[baseline=(char.base)]{
            \node[shape=circle,draw,inner sep=2pt] (char) {#1};}}
\DeclareMathOperator*{\argmax}{arg \hspace{0.8mm} max}

\begin{document}

\title{Maxwell's demon in a double quantum dot with continuous charge detection}
\date{\today}

\author{Bj\"{o}rn Annby-Andersson}
\email{bjorn.annby-andersson@teorfys.lu.se}
%\affiliation{Physics Department and NanoLund$,$ Lund University$,$ Box 118$,$ 22100 Lund$,$ Sweden.}
 
\author{Peter Samuelsson}
%\affiliation{Physics Department and NanoLund$,$ Lund University$,$ Box 118$,$ 22100 Lund$,$ Sweden.}

\author{Ville F. Maisi}
%\affiliation{NanoLund and Solid State Physics$,$ Lund University$,$ Box 118$,$ 22100 Lund$,$ Sweden.}

\author{Patrick P. Potts}
\affiliation{Physics Department and NanoLund$,$ Lund University$,$ Box 118$,$ 22100 Lund$,$ Sweden.}


\begin{abstract}
Converting information into work has during the last decade gained renewed interest as it gives insight into the relation between information theory and thermodynamics. Here we theoretically investigate an implementation of Maxwell's demon in a double quantum dot and demonstrate how heat can be converted into work using only information. This is accomplished by continuously monitoring the charge state of the quantum dots and transferring electrons against a voltage bias using a feedback scheme. We investigate the electrical work produced by the demon and find a non-Gaussian work distribution. To illustrate the effect of a realistic charge detection scheme, we develop a model taking into account noise as well as a finite delay time, and show that an experimental realization is feasible with present day technology. Depending on the accuracy of the measurement, the system is operated as an implementation of Maxwell's demon or a single-electron pump.
\end{abstract}

\maketitle

\section{Introduction}

The conventional statement of the second law of thermodynamics is formulated for systems on the macroscopic scale and asserts that a change of entropy cannot be negative. It is therefore no surprise that it fails to describe systems that operate on a microscopic level. A feedback-controlled system that violates the conventional formulation of the second law was introduced by James C. Maxwell \citep{1871maxwell,Maxwell-heat-book, physicsofMD-maruyama}, who noted that a creature with the ability of tracking and determining the velocity of individual gas particles could create a temperature gradient. Scientists were confused by \textit{Maxwell's demon} for a long time, but following the work of Bennett \cite{bennett1982} and Landauer \cite{landauer1961}, the process is completely legitimate. The demon acquires and stores information about the gas particles. Deleting this information requires an increase of entropy such that the second law is restored. These works established the close relationship between thermodynamics and information theory, suggesting that for feedback-controlled systems, the role of information must be incorporated into the formulation of the second law.

The field of stochastic thermodynamics deals with the microscopic nature of thermodynamics \cite{review3,review8,review4,Horowitz-NatPhys-2020}. It has revealed several important results, e.g.~fluctuation theorems which generalize the second law to the nano-scale. During the course of the last decade, the incorporation of information in stochastic thermodynamics has resulted in a number of novel insights \cite{2002maxwell-demon-2,sagawa-info-thermo,physicsofMD-maruyama,parrondo2015thermo-of-info,PatrickPeter-TUR}. In particular, fluctuation theorems have been generalized to hold for feedback-controlled systems \cite{sagawa-ueda-2008secondlaw,info-fluct-theorem-1,nonequilibrium-feedback-control,info-fluct-theorem-2,info-fluct-theorem-3,info-fluct-theorem-4,PatrickPeter-measurement-feedback} showing, in agreement with Bennett's insight, that information can give rise to negative entropy production, and needs to be taken into account in the thermodynamic bookkeeping. Due to experimental advances, it is nowadays possible to control systems down to the nano-scale, making it possible to realize Maxwell's demon in the lab. Several experimental studies have been conducted, using Brownian particles \cite{brownianparticleMD}, molecular machines \cite{molecularmachineMD}, photonic systems \cite{photonicMD}, electronic systems \cite{koskiSzilard2014-1,koskiSzilard2014-2,SEB-MD-chida,koski2015chip-MD}, ultracold atoms \cite{ultracold-atoms-demon}, and DNA molecules \cite{DNA-demon}. In addition, several experiments on Maxwell's demon in the quantum regime have recently been presented \cite{quantumMD-1,quantumMD-2,quantumMD-3}.

An advancement that is particularly promising for investigating the thermodynamics of information is provided by single electron charge detection in quantum dots \cite{single-electron-gustavsson,SET-1,SET-2,QPC-1,QPC-2,single-electron-fujisawa}. Detection of single electrons becomes highly important when realizing Maxwell's demon in electronic circuits, where the electrons resemble the gas particles in the original thought experiment. Common charge detectors involve single electron transistors \cite{SET-1,SET-2} and quantum point contacts (QPC) \cite{QPC-1,QPC-2}. Such detectors are coupled capacitively to the electronic circuit under investigation. If the current through the detector depends sensitively on nearby charges, individual tunneling events of electrons can be resolved in real time. Experimental implementations of Maxwell's demon based on charge sensing in single electron boxes are given in Refs.~\cite{koskiSzilard2014-1,koskiSzilard2014-2,SEB-MD-chida}.

In this paper, we investigate an implementation of Maxwell's demon in an electronic system constituted by two single-level quantum dots coupled to electron reservoirs at the same temperature. We theoretically demonstrate how heat can be converted into work by transferring electrons against an external voltage bias using a measurement and feedback scheme. During the process, the electron occupation of the dots is measured, and this information is used to guide the electrons through the system by tuning the energy level positions of the dots. We provide a quantitative statistical analysis of the electron transport by employing full counting statistics (FCS). From the FCS, the statistics of work and heat can directly be extracted. We first consider an ideal detector providing perfect knowledge on the charge state at each moment in time. Then we investigate a realistic detector where both delay and noise are taken into account.

An analogous operation cycle was first presented by Averin, M\"{o}tt\"{o}nen, and Pekola in Ref.~\cite{Pekola2011MD-theory} for a system constituted by metallic islands. In this paper, we go beyond their results by considering a complete statistical study on the electron transport, and an implementation of a realistic detector model. Furthermore, in contrast to Refs.~\cite{koskiSzilard2014-1,koskiSzilard2014-2,Pekola2011MD-theory}, our implementation is based on quantum dots, offering the possibility of tuning electron tunneling rates. Furthermore, the well defined energy levels of quantum dots allow for accessing heat fluctuations. We note that it is easy to establish a large inter-dot Coulomb blockade in the double dot system which may provide a challenge in metallic islands.

Previously proposed implementations of Maxwell's demon based on charge sensing in a single quantum dot have been presented in Refs.~\cite{schaller2011MD-theory,Schaller-PRB-2018}, where electrons are transferred against a voltage bias by tuning electron tunneling rates. Another class of implementations are autonomous Maxwell demons, which do not rely on an external measurement and feedback loop. Several theoretical \cite{schaller2013MD-theory,Kutvonen-SciRep-2016,Strasberg-PRB-2018,Erdman-PRB-2018,Patrick-auto-demon} as well as experimental \cite{koski2015chip-MD} studies making use of autonomous charge sensing have been presented.

The paper is organized as follows. In Sec.~\ref{sec:system-operation}, the double quantum dot system is introduced together with a description of the ideal operation cycle. Section \ref{sec:ideal-demon} is devoted to the results obtained when operating the system ideally, while Sec.~\ref{sec:NON-ideal-demon} presents the results for a realistic detector.

\section{System and ideal operation}
\label{sec:system-operation}

\begin{figure}[H]
\centering
\includegraphics[width=1\columnwidth]{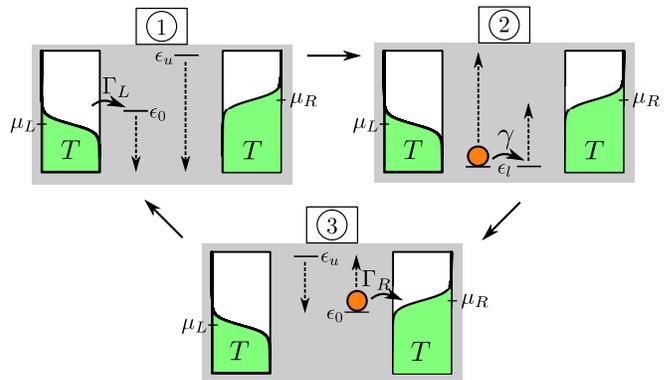}
\caption{Visualization of the demon cycle. Curved arrows depict possible electron tunneling events, where $\Gamma_L$, $\Gamma_R$, and $\gamma$ are electron tunneling rates. As soon as such an event occurs, it is immediately detected and the energy levels are moved as illustrated by the dashed arrows.}
\label{fig:demon-cycle}
\end{figure}

By confining electrons in a small volume, for instance in a nanowire, it is possible to define an atomlike structure that is known as a quantum dot. By monitoring the charge states of these structures, and modifying their internal properties, it is possible to implement Maxwell's demon. Here, we consider two quantum dots coupled in series to each other with tunneling rate $\gamma$, and to an equilibrium electron reservoir each with tunneling rates $\Gamma_L$ and $\Gamma_R$, respectively. The reservoirs $L$ and $R$ are described by the Fermi-Dirac distribution 
\begin{equation}
f_{L/R}(\epsilon) = \big[ e^{(\epsilon-\mu_{L/R})/k_BT} + 1 \big]^{-1},
\label{eq:fermi-functions}
\end{equation}
\noindent where $\epsilon$ denotes energy, $T$ the temperature, and $\mu_{L/R}$ the chemical potential. We assume large intra- and inter-dot Coulomb repulsion such that only one electron can reside in the double quantum dot, and we neglect degeneracies (e.g.~spin). Finally, we assume that the coherence time is shorter than any other relevant time-scale such that superpositions of charge states can be neglected. This implies that the system can be in one of three distinct states: $(0,0)$, $(1,0)$, and $(0,1)$, corresponding to empty dots, left dot occupied, and right dot occupied, respectively. We assume that the energy levels of the dots, denoted by $\{ \epsilon_L,\epsilon_R \}$, can be tuned by external voltage gates. We restrict ourselves to the three level settings: $\{\epsilon_0,\epsilon_u\}$, $\{\epsilon_l,\epsilon_l\}$, and $\{\epsilon_u,\epsilon_0\}$, illustrated in Fig.~\ref{fig:demon-cycle}.

Our aim is to show that electrons can be moved against an external voltage bias $\mu_R-\mu_L=eV$ without performing work by the voltage gates. This is achieved by measuring the occupation of the dots and applying feedback. The operation we consider is illustrated in Fig.~\ref{fig:demon-cycle} and shows how the electrons are transferred against the bias using information alone. In this and the next section, we consider ideal operation conditions. This corresponds to three assumptions. First, we assume that the continuous occupation measurements of the dots are error-free such that we know the system state with certainty at all times. Second, it is assumed that the feedback is applied instantaneously. Finally, we choose $\mu_{L/R} - \epsilon_l \gg k_BT$ and $\epsilon_u - \mu_{L/R} \gg k_BT$ such that the occupation probabilities for $\epsilon_l$ and $\epsilon_u$ can be approximated as $f_{L/R}(\epsilon_l)=1$ and $f_{L/R}(\epsilon_u)=0$. In Sec.~\ref{sec:NON-ideal-demon}, we will consider a non-ideal demon, where the first and third assumptions will be relaxed. Letting $\mu_L < \mu_R$, the following cyclic scheme can be utilized:

\begin{enumerate}
\item Initially, the dots are empty, and the energy levels are set to $\{ \epsilon_0,\epsilon_u \}$, as visualized in \circled{1}, see Fig.~\ref{fig:demon-cycle}. With these settings, the only event possible is an electron tunneling into the left dot. When this occurs, the levels are immediately moved to $\{ \epsilon_l , \epsilon_l \}$, see \circled{2}, such that the electron cannot tunnel back to the left reservoir.

\item As the electron tunnels to the right dot, the energy levels are moved instantly to $\{ \epsilon_u , \epsilon_0 \}$, as depicted in $\circled{3}$.

\item The electron can now only tunnel to the right reservoir. As this happens, the levels are moved back to their initial position in \circled{1}, closing the cycle.
\end{enumerate}

%\begin{figure}[H]
%\centering
%\includegraphics[width=\columnwidth]{cycle1.eps}
%\caption{The demon prepares the system as depicted in 1 with empty dots. As an electron tunnels into the left dot, the demon moves the energy levels as visualized in 2. When the electron tunnels to the right dot, the levels are shifted as depicted in 3. Just as the electron leaves the right dot and enter the right reservoir, the demon brings the system back to step 1 and the cycle is complete. Solid arrows depict possible electron tunneling events, where $\Gamma_L$, $\Gamma_R$, and $\gamma$ are electron tunneling rates. As soon as such an event occurs, it is immediately detected and the energy levels are moved as illustrated by the dashed arrows.}
%\label{fig:demon-cycle}
%\end{figure}
%
For each cycle, the first law of thermodynamics can be written as $Q + W + W_g = 0$. The work $W = \mu_R - \mu_L = eV$ is performed by moving one electron from a region with lower to a region with higher chemical potential. The corresponding heat $Q_L = \mu_L - \epsilon_0$ and $Q_R = \epsilon_0 - \mu_R$ enters the left and right reservoirs, cooling both reservoirs when $\mu_L < \epsilon_0 < \mu_R$. The total heat is given by $Q = Q_L + Q_R = - W$. The total work extracted by the gates $W_g = (\epsilon_0 - \epsilon_l) + (\epsilon_l - \epsilon_0) = 0$ vanishes, as desired. Finally, the change in system energy is zero as the process is cyclic. Here, we use the convention $Q>0$ if energy is entering a reservoir, and $W>0$ when work is produced. Thus, for $eV \geq 0$, heat is converted into work using information.

\section{Ideal demon: Quantitative description}
\label{sec:ideal-demon}

Under ideal conditions, the occupation of the quantum dots uniquely determines the level positions $\{\epsilon_L,\epsilon_R\}$. Hence, the system dynamics can be described by the rate equation $\boldsymbol{\dot{\rho}}_0(t) = \mathcal{L}_0 \boldsymbol{\rho}_0(t)$, where the components $\rho_{0,j}(t)$, $j= (0,0)$, $(1,0)$, $(0,1)$, of $\boldsymbol{\rho}_0(t)$ are the probabilities to occupy state $j$ at time $t$. The Liouvillian $\mathcal{L}_0$ and the transition rates between the system states are determined by the following transitions representing the ideal cycle
\begin{equation}
(0,0) \xrightarrow{\makebox[0.3cm]{$\gamma_L$}} (1,0) \xrightarrow{\makebox[0.3cm]{$\gamma$}} (0,1) \xrightarrow{\makebox[0.3cm]{$\gamma_R$}} (0,0),
\label{eq:cycle}
\end{equation}
\noindent where
\begin{equation}
\gamma_L = \Gamma_L f_L(\epsilon_0), \hspace{10mm} \gamma_R = \Gamma_R \left[ 1 - f_R(\epsilon_0) \right], 
\label{eq:transition_rates}
\end{equation}
\noindent see Fig.~\ref{fig:demon-cycle}.

\subsection{Full counting statistics}

To describe the statistics of the electron transport in the system, we use FCS \citep{levitov-96-FCS}, the central object of which is given by the probability distribution of the number of electrons transferred to the right reservoir within the time $t$, $p(n,t) = \sum_{j} \rho_j(n,t)$, with $\rho_j(n,t)$  being the components of the number-resolved probability vector $\boldsymbol{\rho}(n,t)$. With the discrete Fourier transform of $\boldsymbol{\rho}(n,t)$ given by $\boldsymbol{\rho}(\chi,t)$ where the counting field $\chi$ is introduced, the dynamics of the system is governed by the differential equation \cite{schaller-book} 
\begin{equation}
\dot{\boldsymbol{\rho}}(\chi,t) = \mathcal{L}(\chi) \boldsymbol{\rho}(\chi,t),
\label{eq:countingfield_rate_eq}
\end{equation}
\noindent with the Liouvillian 
\begin{equation}
\mathcal{L}(\chi) =
\begin{pmatrix}
-\gamma_L & 0 & e^{i\chi}\gamma_R \\
\gamma_L & -\gamma & 0 \\
0 & \gamma & -\gamma_R
\end{pmatrix}
.
\label{eq:ideal_liouvillian}
\end{equation}
\noindent By putting $\chi = 0$, the rate equation in Eq.~(\ref{eq:cycle}) is recovered, i.e. $\mathcal{L}_0 = \mathcal{L}(0)$.

The cumulant generating function (CGF) $\mathcal{C}(\chi,t)$ of the probability distribution is defined via
\begin{equation}
e^{\mathcal{C}(\chi,t)} = \sum_{n} p(n,t) e^{in\chi} = \sum_{j} \rho_j (\chi,t),
\label{eq:cumulant-generating-function}
\end{equation}
\noindent and the cumulants are obtained using
\begin{equation}
\langle \! \langle n^k(t) \rangle \! \rangle = (-i)^k \left. \frac{\partial ^k}{\partial \chi^k} \mathcal{C}(\chi,t) \right| _{\chi = 0}.
\label{eq:cumulant-formula}
\end{equation}
\noindent In the long time-limit, the CGF is given, up to a time independent correction term which will be dropped in the following, by $\mathcal{C}(\chi,t) \approx \lambda (\chi) t$, where $\lambda(\chi)$ is the eigenvalue of the Liouvillian that satisfies $\lambda(0) = 0$ \cite{schaller-book}. The eigenvalue is given by one of the roots of the characteristic polynomial of $\mathcal{L}(\chi)$
%
%\begin{equation}
%\lambda^3 + \xi \lambda^2 + \kappa \lambda + \eta (1-e^{i\chi}) = 0,
%\label{eq:characteristic_polynimial_liouvillian}
%\end{equation}
%
\begin{equation}
\begin{aligned}
\lambda ^3 + (\gamma + \gamma_L + \gamma_R) \lambda^2 + (&\gamma \gamma_L + \gamma \gamma_R + \gamma_L \gamma_R) \lambda \\ &+ \gamma \gamma_L \gamma_R (1-e^{i \chi}) = 0.
\end{aligned}
\label{eq:characteristic_polynimial_liouvillian}
\end{equation}
%
%\noindent where,
%
%\begin{equation}
%\begin{aligned}
%\xi &= \gamma + \gamma_L + \gamma_R, \\
%\kappa &= \gamma \gamma_L + \gamma \gamma_R + \gamma_L \gamma_R, \\
%\eta &= \gamma \gamma_L \gamma_R .
%\end{aligned}
%\label{eq:coefficients-charac-eq}
%\end{equation}

\noindent The solutions of this equation do not provide additional physical insight, and are therefore not given here. Note that Eq.~(\ref{eq:characteristic_polynimial_liouvillian}) is invariant under the exchange of any two transition rates, implying that also the cumulants possess this symmetry. The first, second, and third order cumulants are given by 

%\begin{subequations}
%\begin{align}
%\langle \! \langle n(t) \rangle \! \rangle &= \frac{\eta}{\kappa} t, \\
%\langle \! \langle n^2(t) \rangle \! \rangle &= \eta \frac{\kappa^2 - 2 \eta \xi}{\kappa^3} t, \\
%\langle \! \langle n^3(t) \rangle \! \rangle &= \eta \frac{\kappa^4 - 6 \eta^2 \kappa -6 \eta \xi(\kappa^2 - 2 \eta \xi)}{\kappa ^5} t .
%\end{align}
%\label{eq:1st_cumulant}
%\end{subequations}
%
\begin{equation}
\begin{aligned}
\langle \! \langle n(t) \rangle \! \rangle &= \frac{\gamma \gamma_L \gamma_R}{\gamma \gamma_L \! + \! \gamma \gamma_R \! + \! \gamma_L \gamma_R} t, \\ 
%NEXT EQUATION
\langle \! \langle n^2(t) \rangle \! \rangle &= \langle \! \langle n(t) \rangle \! \rangle \frac{\gamma^2 \gamma_L^2 + \gamma^2 \gamma_R^2 + \gamma_L^2 \gamma_R^2}{\left(\gamma \gamma_L + \gamma \gamma_R + \gamma_L \gamma_R \right)^2}, \\
%NEXT EQUATION
\langle \! \langle n^3 (t) \rangle \! \rangle &= \langle \! \langle n(t) \rangle \! \rangle  \left[ 1 - \frac{6 \langle \! \langle n(t) \rangle \! \rangle}{\gamma \gamma_L \gamma_R t } \right. \\ & \times \left. \left( \frac{\langle \! \langle n(t) \rangle \! \rangle ^2}{t^2} - (\gamma \! + \! \gamma_L \! + \! \gamma_R) \frac{\langle \! \langle n^2 (t) \rangle \! \rangle}{t} \right) \right].
\end{aligned}
\label{eq:1st_cumulant}
\end{equation}
Finally, the full distribution is obtained through
\begin{equation}
p(n,t) = \frac{1}{2\pi}  \int ^{2\pi}_{0} d\chi e^{-in\chi + \lambda (\chi)t}.
\label{eq:distribution-steady-state}
\end{equation}
\subsection{Power production}

\begin{figure}[H]
\centering
\includegraphics[width=\columnwidth]{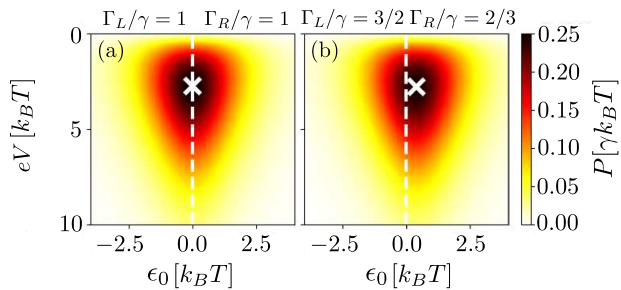}
\caption{First order power cumulant $\langle \! \langle P \rangle \! \rangle$ in units of $\gamma k_BT$ as a function of $\epsilon_0$ and the voltage bias $eV$ for tunneling rates (a) $\Gamma_L/\gamma=\Gamma_R/\gamma=1$, and (b) $\Gamma_L/\gamma=3/2$, $\Gamma_R/\gamma=2/3$. The white, dashed line emphasizes where $\epsilon_0 = 0$, and the white cross indicates where the power reaches its maximum. Here, the chemical potentials are chosen as $\mu_R=eV/2=-\mu_L$ which is not necessarily symmetric around $\epsilon_0$. For non-equal tunneling rates, a symmetric bias around $\epsilon_0$ is no longer optimal.}
\label{fig:2D-power-panel}
\end{figure}

We are interested in the electrical power $P=V \cdot I$ which is a measure of the heat-to-work conversion resulting from the demon operation. Note that a positive current here is defined to flow against the bias. The power is related to the produced work via $W = \int P dt$. The low frequency power cumulants for the system studied here are given by
\begin{equation}
\langle \! \langle P^k \rangle \! \rangle = (eV)^k \lim_{t\rightarrow \infty} \frac{\langle \! \langle n^k(t) \rangle \! \rangle}{t}  .
\label{eq:power-cumulants}
\end{equation}
We start by considering the mean power
\begin{equation}
\frac{\langle \! \langle P \rangle \! \rangle}{\gamma k_B T} = \frac{eV}{k_B T} \frac{\gamma_L \gamma_R}{\gamma \gamma_L + \gamma \gamma_R + \gamma_L \gamma_R},
\label{eq:1st-power-cumulant}
\end{equation}
\noindent that is plotted in Fig.~\ref{fig:2D-power-panel} as a function of $\epsilon_0$ (cf.~Fig.~\ref{fig:demon-cycle}) and the voltage bias $eV$. The plots show that $\langle \! \langle P \rangle \! \rangle \geq 0$ for all parameters which is due to the ideal operation conditions ensuring that electrons are only transferred from left to right. This demonstrates that information can be used to convert heat into work in this system. From Eq.~(\ref{eq:1st-power-cumulant}), it is clear that the power increases with increasing tunneling rates. For fixed tunneling rates, $\Gamma_L$, $\Gamma_R$, $\gamma$, the power attains its largest value at
\begin{subequations}
\begin{align}
\frac{\epsilon_0}{k_BT} &= \frac{1}{2}\ln \left[ \frac{\Gamma_L}{\Gamma_R} \right], \\
\frac{eV}{k_BT} &= 2 \left[ W \left( \frac{\Gamma_L + \Gamma_R + \Gamma_L \Gamma_R / \gamma}{2 e \sqrt{\Gamma_L \Gamma_R}} \right) + 1 \right], 
\end{align}
\label{eq:maxumum_power_position}
\end{subequations}
\noindent where $W(z)$ is the Lambert W-function \cite{lambert-w-paper}, also known as the product logarithm function. Note that $e$ on the right hand side of Eq.~(\ref{eq:maxumum_power_position}b) is Euler's number and not the elementary charge. For $\Gamma_L = \Gamma_R$, the maximum is located at $\epsilon_0 = 0$, i.e.~in the middle of the bias window. For $\Gamma_L \neq \Gamma_R$, this is no longer true; compare Figs.~\ref{fig:2D-power-panel}(a) and (b). 

Another feature in Fig.~\ref{fig:2D-power-panel}(a), where $\Gamma_L=\Gamma_R$, is the mirror symmetry around $\epsilon_0 = 0$. This corresponds to the transformation $\gamma_L \leftrightarrow \gamma_R$, which is the exchange-symmetry of transition rates in the charactersitic polynomial discussed above. In Fig.~\ref{fig:2D-power-panel}(b), where $\Gamma_L \neq \Gamma_R$, the transformation $\gamma_L \leftrightarrow \gamma_R$ has a more complicated effect.
\subsection{Fano factor and skewness}

The Fano factor is defined as the ratio between the second and first order cumulants [cf.~Eq.~(\ref{eq:cumulant-formula})], and gives a measure on the statistical character of the electron transport. For a Poisson process, the Fano factor is 1. A smaller Fano factor implies sub-, a larger super-Poissonian statistics corresponding to anti-bunching \cite{Wagner-NatNanotechnol-2017} and bunching \cite{Maisi-PRL-2014,Maisi-PRL-2016,Fujita-PRL-2016} of electrons, respectively.

\begin{figure}[H]
\centering
\includegraphics[width=1\columnwidth]{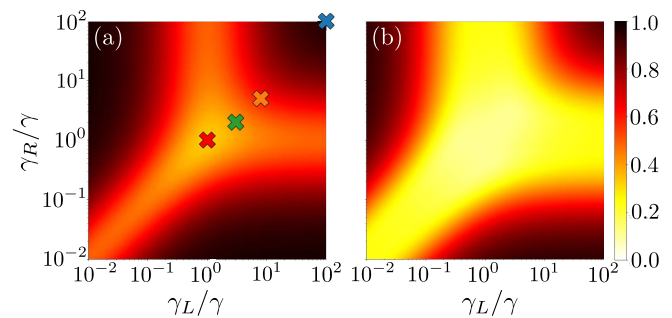}
\caption{\label{fig1} (a) Fano factor $\langle \! \langle n^2(t) \rangle \! \rangle / \langle \! \langle n(t) \rangle \! \rangle$ as function of $\gamma_L/\gamma$ and $\gamma_R/\gamma$. The Fano factor is bounded from above by 1, and from below by 1/3. The upper bound is obtained when one transition rate is much smaller than the others, corresponding to the dark areas in the corners. The lower bound is obtained when all transition rates are equal, corresponding to the bright area in the middle. The colored crosses visualize the choice of parameters in Fig.~\ref{fig:distributions}. (b) Normalized third order cumulant $\langle \! \langle n^3(t) \rangle \! \rangle / \langle \! \langle n(t) \rangle \! \rangle$. This factor only assumes finite values implying non-Gaussian statistics for work and heat.}
\label{fig:fano-factor}
\end{figure}

With the first and second order cumulants given by Eqs.~(\ref{eq:1st_cumulant}), the Fano factor reads,
\begin{equation}
\frac{1}{eV} \frac{\langle \! \langle P^2 \rangle \! \rangle}{\langle \! \langle P \rangle \! \rangle} = \frac{\langle \! \langle n^2(t) \rangle \! \rangle}{\langle \! \langle n(t) \rangle \! \rangle} = \frac{\gamma^2 \gamma_L^2 + \gamma^2 \gamma_R^2 + \gamma_L^2 \gamma_R^2}{\left(\gamma \gamma_L + \gamma \gamma_R + \gamma_L \gamma_R \right)^2}.
\label{eq:fano_factor}
\end{equation}
\noindent The Fano factor is visualized in Fig.~\ref{fig:fano-factor}(a) as a function of the weighted transition rates $\gamma_L/\gamma$ and $\gamma_R/\gamma$. The symmetry of exchanging $\gamma_L$ and $\gamma_R$ discussed above is evident in the figure. In our system, we find the Fano factor to be bounded by
\begin{equation}
\frac{1}{3} \leq \frac{\langle \! \langle n^2(t) \rangle \! \rangle}{\langle \! \langle n(t) \rangle \! \rangle} \leq 1,
\label{eq:fano-factor-inequalities}
\end{equation}
\noindent where the lower bound corresponds to having all transition rates equal. A Fano factor smaller than one is expected since only a single electron can occupy the system at a time resulting in anti-bunching. The upper bound occurs when any of the transition rates is much smaller than the other two, e.g.~when $\gamma \ll \gamma_L, \gamma_R$. In this regime, the system behaves effectively as a single tunneling barrier and the statistics are Poissonian.

Figure \ref{fig:fano-factor}(b) visualizes the normalized third order cumulant $\langle \! \langle n^3(t) \rangle \! \rangle / \langle \! \langle n(t) \rangle \! \rangle$ which captures the skewness of the distribution $p(n,t)$. Since the third order cumulant is non-zero, the transport statistics, and therefore also the heat and work statistics, are non-Gaussian.  
\subsection{Probability distributions}

The full distribution $p(n,t)$, which is directly related to the work and heat distributions by $W=neV$ and $Q=-W$, can be evaluated in the long time limit using Eq.~(\ref{eq:distribution-steady-state}). It is instructive to consider two limiting cases. First, when one transition rate is much smaller than the others, e.g. $\gamma \ll \gamma_L, \gamma_R$, we may expand the dominant eigenvalue of the Liouvillian up to linear order in $\gamma$, obtaining $\lambda(\chi) = \gamma \left(e^{i\chi} - 1 \right)$. With Eq.~(\ref{eq:distribution-steady-state}), this results in
\begin{equation}
p(n,t) = e^{-\gamma t} \frac{(\gamma t)^n}{n!},
\label{eq:poisson-distribution}
\end{equation}
\noindent which is the Poisson distribution, in agreement with the discussion on the Fano factor above.

\begin{figure}[H]
\centering
\includegraphics[width=0.7\columnwidth]{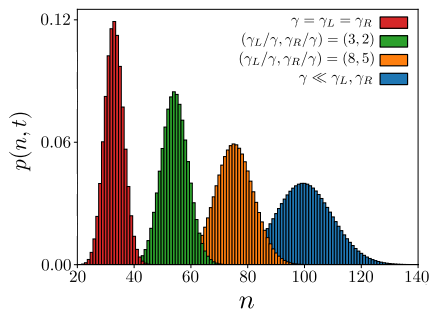}
\caption{Probability distribution $p(n,t)$ in four cases as a function of $n$ for $\gamma t = 100$. The red and blue bars represent the analytical solutions obtained when all transition rates are equal and when one rate is much smaller than the others, respectively [see Eqs.~(\ref{eq:poisson-distribution}) and (\ref{eq:distribution-all-rates-equal})]. The green and orange bars were obtained numerically.}
\label{fig:distributions}
\end{figure}

Second, with all transition rates equal, $\gamma = \gamma_L = \gamma_R$, the total number of tunneling events $q$ in the system is distributed according to Eq.~(\ref{eq:poisson-distribution}). Starting in the state $(0,0)$, an electron must tunnel three times before reaching the right reservoir, see Fig.~\ref{fig:demon-cycle}. To have $n$ electrons transferred to the right reservoir, $q=3n+k$ tunneling events must have occured, with $k=0,1,2$. This results in the distribution
\begin{equation}
p(n,t) = e^{-\gamma t} \sum_{k=0,1,2} \frac{(\gamma t)^{3n+k}}{(3n+k)!}.
\label{eq:distribution-all-rates-equal}
\end{equation}
Equations (\ref{eq:poisson-distribution}) and (\ref{eq:distribution-all-rates-equal}) are visualized in Fig.~\ref{fig:distributions} together with the distributions for $(\gamma_L/\gamma,\gamma_R/\gamma) = (3,2)$ and $(\gamma_L/\gamma,\gamma_R/\gamma) = (8,5)$, calculated numerically. The center of each distribution, i.e. the mean, agrees with the first order cumulant in Eq.~(\ref{eq:1st_cumulant}). As the third order cumulant in Eq.~(\ref{eq:1st_cumulant}) is non-zero and positive [see Fig.~\ref{fig:fano-factor}(b)], the distribution always has a skewness, and is therefore non-Gaussian. The direct relation between $p(n,t)$ and the distributions of work and heat reveals the non-Gaussian statistics of these thermodynamic quantities.

\subsection{Entropy production of the demon}
\label{sec:entrop-demon}
The operation described in Sec.~\ref{sec:system-operation} results in an entropy reduction in the system. It is therefore required that the entropy produced by the demon balances this reduction such that the system and demon together obeys the second law of thermodynamics. Recently, bounds on entropy production in heat engines have been derived within the framework of thermodynamic uncertainty relations (TURs) \cite{Gingrich-PRL-2016,Potts-PRE-2019,Horowitz-NatPhys-2020,Vu-J-of-PhysA-2020}. To derive a TUR for our system and demon is a daunting task, and is not attempted here. However, by assuming that a generic TUR holds, derived for a number of different systems, we can establish a lower bound for the entropy production. The TUR considered is given by
\begin{equation}
\frac{\langle \! \langle I^2 \rangle \! \rangle}{\langle \! \langle I \rangle \! \rangle ^2} \geq \frac{2}{\sigma_S + \sigma_D},
\label{eq:general-TUR}
\end{equation}
\noindent where $\langle \! \langle I \rangle \! \rangle$ and $\langle \! \langle I^2 \rangle \! \rangle$ are the two lowest order current cumulants obtained from $\langle \! \langle I^k \rangle \! \rangle = \lim_{t \rightarrow \infty} e^k \langle \! \langle n^k(t) \rangle \! \rangle / t$ with the elementary charge $e$, and $\sigma_S$ and $\sigma_D$ are the entropy production (in units of $k_B$) for the system and demon, respectively. The entropy production for the system is obtained through $\sigma_S = - \langle \! \langle I \rangle \! \rangle V /k_B T$, which is the heat current into the reservoirs divided by their temperature $T$. For the demon entropy production, we may write $\sigma_D = x_D \langle \! \langle I \rangle \! \rangle / k_B T$, where $x_D$ is the entropy production in the demon in units of $-\sigma_S$. The total entropy production for the system and demon together can then be written as
\begin{equation}
\sigma_S + \sigma_D = \frac{\langle \! \langle I \rangle \! \rangle}{k_B T} \left( x_D -1 \right).
\label{eq:total-entropy}
\end{equation}
\noindent Using this together with the TUR given in Eq.~(\ref{eq:general-TUR}), the following inequality for $x_D$ can be obtained,
\begin{equation}
x_D \geq 1 + \frac{2k_B T}{eV}\frac{1}{F},
\label{eq:bound-demon-entropy}
\end{equation}
\noindent where $F$ is the Fano factor in Eq.~(\ref{eq:fano_factor}). This sets a lower bound on the demon entropy production. As $x_D \geq 1$, it is assured that the total entropy production $\sigma_S+\sigma_D \geq 0$, and thus the second law is respected.

It is instructive to introduce the efficiency
\begin{equation}
\eta \equiv \frac{-\sigma_S}{\sigma_D} = \frac{1}{x_D} \leq \frac{eV \cdot F}{eV \cdot F + 2k_BT},
\label{eq:entropy-efficiency}
\end{equation}
\noindent  which relates the entropy reduction in the system to the entropy produced by the demon. In the reversible limit, where the demon produces the minimal amount of entropy allowed by the second law, the efficiency is maximal and equal to unity. From Eq.~(\ref{eq:entropy-efficiency}), we find that the reversible limit may only be reached for $eV \gg k_B T$.
\section{Non-ideal demon}
\label{sec:NON-ideal-demon}

So far, we have focused on ideal operation conditions relying on an ideal detector that is infinitely fast and error free. Here we consider a more realistic model describing a QPC or quantum dot detector asymmetrically coupled to the double dot system \cite{single-electron-fujisawa,ensslin-doubledots}. In particular, we include delay as well as noise in the charge sensing. These constitute the main error sources in realistic detectors. Furthermore, we relax the requirement that $f_{L/R}[\epsilon_{u(l)}]=0(1)$ for a more realistic description. We note that we still consider the change of the energy levels to happen on a time-scale that is much faster than any other relevant time-scale. This is experimentally feasible \cite{pump1,ensslin-doubledots,koskiSzilard2014-1,koskiSzilard2014-2}.

\subsection{Model}
\label{sec:non-ideal:model}

We start by providing a general formulation of the model for the detector, its output, and the feedback. The system state after time $t$ is denoted by $S(t)$, and assumes the values 0, 1, and 2 representing the charge configurations $(0,0)$, $(1,0)$, and $(0,1)$, respectively. For the detector, we introduce a detector output $d(t)$ which is dependent on the system state $S(t)$, a delay time $\tau_d$, and assume that the detector adds Gaussian white noise with standard deviation $\sigma$ on its output. The energy level setting after time $t$ is described by $D(t)$, also taking the values 0, 1, and 2, but here representing the settings $\{\epsilon_0 , \epsilon_u \}$, $\{\epsilon_l, \epsilon_l \}$, and $\{\epsilon_u , \epsilon_0 \}$, respectively.

The system state $S(t)$ is a stochastic variable whose trajectory is determined by the probability vector $\boldsymbol{\rho}(t)$ that is governed by the rate equation 
\begin{equation}
\boldsymbol{\dot{\rho}}(t) = \mathcal{L}\left[D(t) \right] \boldsymbol{\rho}(t), 
\end{equation}
\noindent where the Liouvillaian $\mathcal{L}\left[D(t) \right]$ is dependent on the current energy level setting. The Liouvillian is given in Appendix \ref{sec:appendix:details-monte-carlo}. $D(t)$ is governed by a measurement-feedback scheme which depends on the detector output $d(t)$. We model the detector output as a random Gaussian variable distributed as
\begin{equation}
p \left[ d(t) \right] = \frac{1}{\sqrt{2\pi \sigma^2}} e^{- \left[ d(t) - m(t) \right]^2 / 2 \sigma^2},
\label{eq:distribution-detector-output}
\end{equation}
\noindent with the standard deviation $\sigma$ quantifying the strength of the noise. The mean of the distribution is given by
\begin{equation}
m(t) = \int_{-\infty}^{t} \tau_d^{-1}e^{-(t-t')/\tau_d} S(t') dt', 
\label{eq:mean-of-d(t)}
\end{equation}
\noindent where the delay $\tau_d$ is introduced. The insets of Fig.~\ref{fig:power-bias-panel} visualize $d(t)$ for some different choices of $\tau_d$ and $\sigma$. With the dots unoccupied, $d(t)$ is distributed around 0, and with one of the dots occupied, the signal is distributed around 1 or 2 depending on the location of the electron. The rise time of the signal is determined by $\tau_d$. The ideal limit corresponds to $\tau_d$, $\sigma \rightarrow 0$, where we find $d(t) = S(t)$ at all times. Having a model for the detector output, we need to provide a feedback protocol, determining the level setting $D(t)$ from the previous output $d(t' \leq t)$. This is provided by
\begin{equation}
D(t) = \argmax _{j \in \{0,1,2\}} \int _{t-4\tau_d}^{t} dt' \theta_j \left[ d(t') \right],
\label{eq:feedback-for-setting}
\end{equation}
\noindent with $\theta_0[x] = \Theta \left[\frac{1}{2}-x\right]$, $\theta_1[x] = \Theta \left[x-\frac{1}{2}\right] \Theta \left[\frac{3}{2}- x\right]$, and $\theta_2[x] = \Theta [x-\frac{3}{2}]$, where $\Theta [x]$ is the Heaviside step function. Equation (\ref{eq:feedback-for-setting}) works as follows; if $d(t') < \frac{1}{2}$ in the majority of the interval $[t-4\tau_d,t]$, below referred to as checking interval, the charge configuration is assumed to be $(0,0)$, i.e.~$S(t)=0$, and the level setting is put to $D(t)=0$. When $\frac{1}{2} \leq d(t') < \frac{3}{2}$ in the majority of the checking interval, the system is assumed to occupy $S(t)=1$, and the level setting $D(t)=1$ is chosen. Finally, as $d(t') \geq \frac{3}{2}$ in the majority of the checking interval, the level setting is put to $D(t)=2$. In contrast to standard threshold detection \cite{ensslin-doubledots,single-electron-koski}, the use of a checking interval reduces the risk of misinterpreting transitions. For instance, this ensures that the level position setting does not pass through $\{ \epsilon_l, \epsilon_l \}$ ($D(t)=1$) when $d(t')$ changes from 2 to 0 in a short time. For $\tau_d \rightarrow 0$, the checking interval becomes infinitely small, and $D(t)$ is determined by the instantaneous value of $d(t)$, leading to $D(t)=d(t)=S(t)$ in the ideal limit, where $\sigma \rightarrow 0$. The feedback is thus infinitely fast in the ideal limit, and we recover the results of the previous section.

\subsection{Monte Carlo simulations}

\begin{figure*}
\centering
\includegraphics[width=0.95\textwidth]{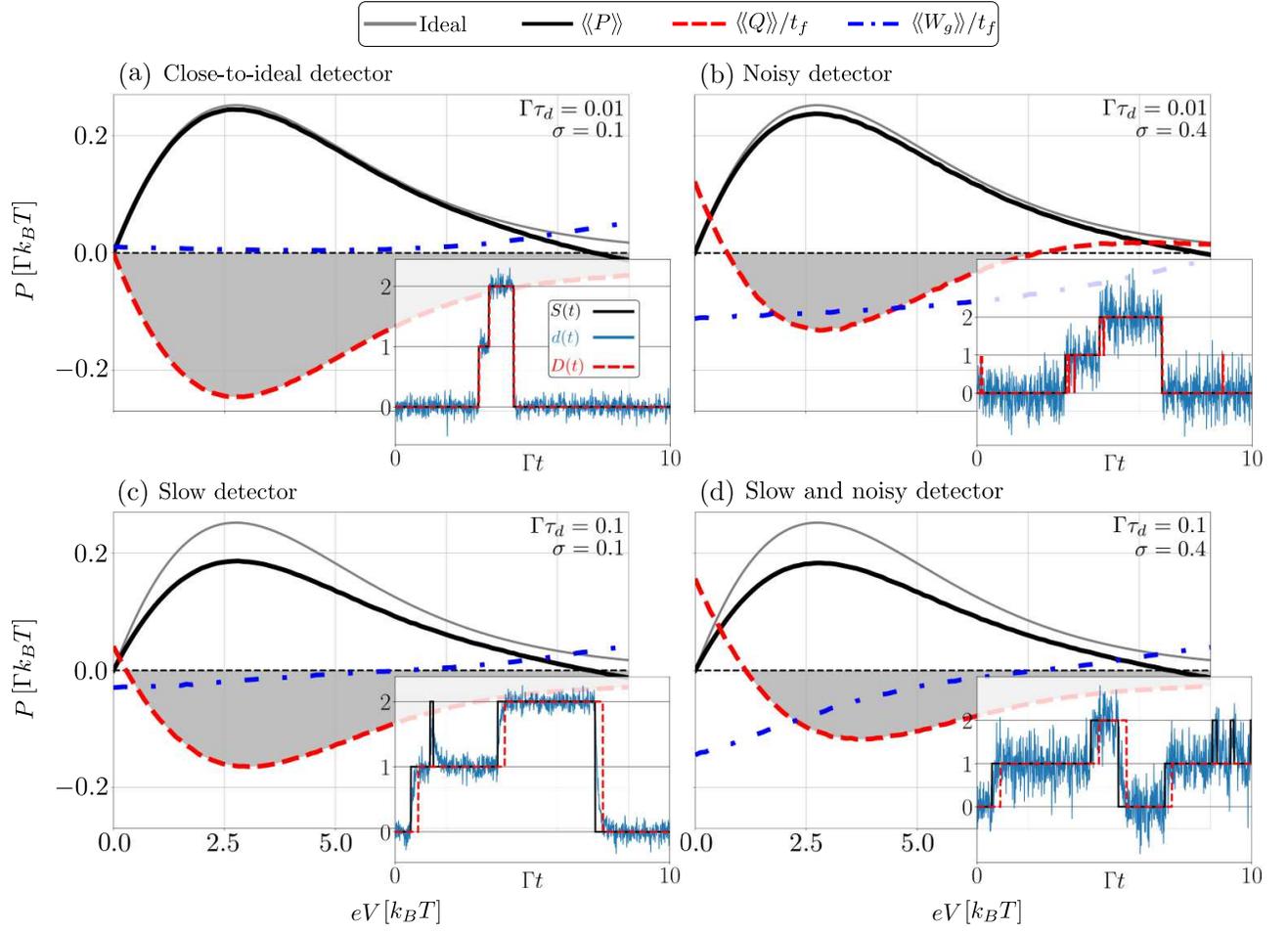}
\caption{Energy fluxes as functions of the voltage bias. (a)-(d) visualize the effect of different choices for $\tau_d$ and $\sigma$. In the gray, shaded regions, the heat is negative, seemingly violating the second law, and the system is operated as a Maxwell demon. The insets show examples of time traces of the true system state $S(t)$, the detector output $d(t)$, and the energy level setting $D(t)$. The thin, gray line is the average power obtained for ideal operation conditions, cf.~Eq.~(\ref{eq:1st-power-cumulant}). (a) Close-to-ideal case. At large voltages, the rates $\gamma_L, \gamma_R$ become small such that the long time limit is no longer reached. (b) Noisy detector. Errors induced by noise result in heating by the voltage gates. (c) Slow detector. The power output is reduced as the detector cannot follow all the tunneling events. (d) Slow and noisy detector. The delay reduces the errors introduced by noise through the averaging over the checking interval $[t-4\tau_d,t]$. For these simulations, we used $\Gamma = \gamma = \Gamma_L = \Gamma_R$, $\epsilon_u = 10k_BT = - \epsilon_l$, $\epsilon_0 = 0$, and $\Gamma t_f=200$.}
\label{fig:power-bias-panel}
\end{figure*}

To simulate the non-ideal scenario, we use a Monte Carlo method to capture the stochastic dynamics of the system. By generating a large number of time traces for $S(t)$, $d(t)$, and $D(t)$, the average work and heat production can be extracted. Details concerning the Monte Carlo method are given in Appendix \ref{sec:appendix:details-monte-carlo}. In Fig.~\ref{fig:power-bias-panel}, we present the main results where the average energy fluxes $\langle \! \langle P \rangle \! \rangle $, $\langle \! \langle Q \rangle \! \rangle /t_f$, and $\langle \! \langle W_g \rangle \! \rangle / t_f$ are plotted. Here, $t_f$ is the total operation time, and $\langle \! \langle \cdot \rangle \! \rangle$ denotes the sample mean of 10 000 Monte Carlo simulations.

\textit{Close-to-ideal detector.} Figure \ref{fig:power-bias-panel}(a) visualizes the results for an almost ideal detector. The grey shaded region indicates where the heat $\langle \! \langle Q \rangle \! \rangle$ is negative. In this region, the second law of thermodynamics is seemingly violated as in Maxwell's original thought experiment. The produced power $\langle \! \langle P \rangle \! \rangle$ recovers the ideal power production for small voltage bias. For large bias, the power becomes negative, $\langle \! \langle P \rangle \! \rangle < 0$. In this regime, the transition rates become very small such that the long time limit may not be reached. With large bias, it becomes more probable to occupy (0,1) in level setting $\{\epsilon_0, \epsilon_u\}$. If the level setting then is changed to $\{\epsilon_u, \epsilon_0\}$, the system gets stuck in the latter level position with the right dot occupied. With this process the gates are cooling the right reservoir and we find $\langle \! \langle Q \rangle \! \rangle <0 $ and $\langle \! \langle W_g \rangle \! \rangle > 0$. The small deviations from $\langle \! \langle W_g \rangle \! \rangle = 0$ for small bias are due to boundary effects such as interrupting the operation cycle at the end of a simulation run.

\textit{Noisy detector.} In Fig.~\ref{fig:power-bias-panel}(b), we present the results for a noisy detector. When increasing $\sigma$, the heat $\langle \! \langle Q \rangle \! \rangle$ becomes positive for small and large bias. In these regions, the second law holds even when disregarding the demon. For large $\sigma$, $D(t)$ starts to fluctuate, see the inset of (b). These fluctuations lead to mistakes in the operation of the system. As $\langle \! \langle W_g \rangle \! \rangle < 0$, the gates are performing work on the system, resulting in the reservoirs being heated. A possible trajectory capturing this is the following
\begin{center}
\begin{tikzpicture}
\node [draw=none, fill=none] (init) {$(0,0) \colon \{\epsilon_0,\epsilon_u\}$};
\node [draw=none, fill=none=init, below of=init, node distance=1.0cm] (sec-state) {$(1,0) \colon \{\epsilon_l,\epsilon_l\}$};
\node [draw=none, fill=none, right of=sec-state, node distance=2.8cm] (third-state) {$(1,0) \colon \{\epsilon_u,\epsilon_0\}$};
\node [draw=none, fill=none, above of=third-state, node distance=1.0cm] (fourth-state) {$(0,0) \colon \{\epsilon_u,\epsilon_0\}$};
%Arrows
\draw [->] (init) -- (sec-state); 
\draw [->,dashed,red,thick] (sec-state) -- (third-state);
\draw [->] (third-state) -- (fourth-state);
\draw [->] (fourth-state) -- (init);
\end{tikzpicture}
\end{center}
\noindent where the dashed, red arrow indicates where the mistake happens; the energy level setting is changed to $\{\epsilon_u,\epsilon_0\}$ even though the electron has not tunneled to the right dot.

\textit{Slow detector.} The results obtained for a slow detector is given in Fig.~\ref{fig:power-bias-panel}(c). When increasing $\tau_d$, the produced power $\langle \! \langle P \rangle \! \rangle$ decreases compared to the ideal curve. The main mechanism behind this is back-tunneling of electrons: Before the detector has registered that a tunneling event occurred, the electron may tunnel back. Such an event is depicted in the inset of (c). For usual experimental operation conditions, here considered to be $\Gamma \tau_d = 0.1$, it is not expected to recover the power that is produced under ideal conditions.

\textit{Slow and noisy detector.} Figure \ref{fig:power-bias-panel}(d) shows the results for a detector which is both slow and noisy. Noteworthy is that the heat for large bias is negative. This is a result of reducing the number of mistakes in the operation compared to the noisy detector, cf.~(b). Mistakes due to noise are reduced for finite $\tau_d$ as the noise is averaged over the checking interval $[t-4\tau_d,t]$, see Eq.~(\ref{eq:feedback-for-setting}).

\begin{figure}
\centering
\includegraphics[width=0.95\columnwidth, angle=0]{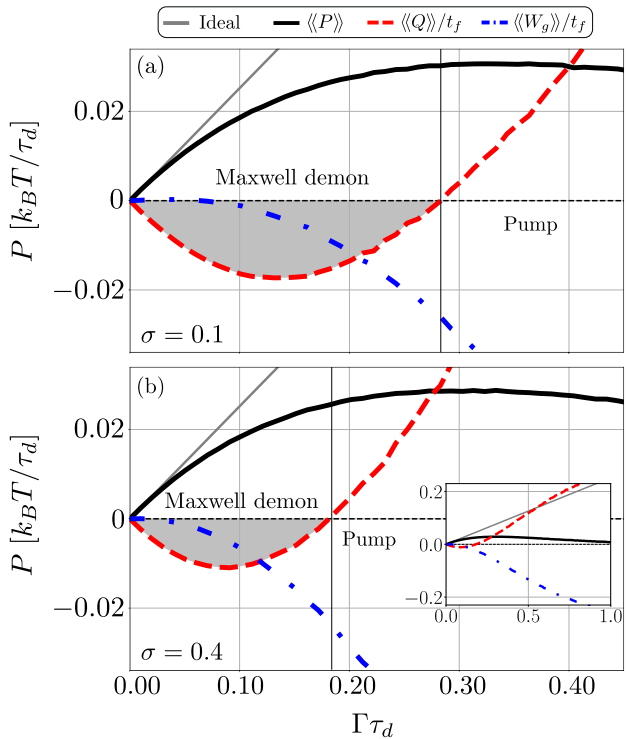}
\caption{\label{powergamma2} Energy fluxes as functions of $\Gamma \tau_d$ where $\Gamma$ is varied and $\tau_d$ is fixed. The noise is chosen as $\sigma=0.1$ and $\sigma=0.4$ in (a) and (b), respectively. The system is operated as a Maxwell demon in the gray, shaded regions as the heat is negative there. This is indicated to the left of the thin, black vertical line. To the right of this line, the system is behaving like an electron pump, where the gate voltages ($W_g$) are used to transfer electrons against the voltage bias. The ideal case [i.e.~Eq.~(\ref{eq:1st-power-cumulant})] is represented by the thin, gray line. The inset in (b) shows a zoom-out of the plot. For each point in the plots, $\Gamma t_f = 100$, and $eV/k_BT = 2.8$ which roughly corresponds to the maximum in the ideal case, Eq.~(\ref{eq:maxumum_power_position}b).}
\label{fig:power-gamma}
\end{figure}     
Figure \ref{fig:power-gamma} visualizes the same energy fluxes as Fig.~\ref{fig:power-bias-panel}, but as functions of $\Gamma \tau_d$ keeping $\tau_d$ fixed. We present two cases with different noise levels. We see how the tunneling rate affects the ability to operate the system in accordance with the ideal cycle. The ideal case is recovered for $\Gamma \tau_d \lesssim 0.05$. In this regime, the detector can resolve the electron trajectories. Thereby, we can afford to slightly increase $\Gamma$ and observe an increase of $\langle \! \langle P \rangle \! \rangle$ because of a larger number of transferred electrons per unit time. Increasing $\Gamma$ further, the detector can no longer resolve the trajectories. This trade-off results in a non-monotonic behavior of the power. When $\Gamma \rightarrow \infty$, $\langle \! \langle P \rangle \! \rangle \rightarrow 0 $ as can be seen in the inset of Fig.~\ref{fig:power-gamma}(b). From Fig.~\ref{fig:power-gamma}(a), we infer that detector delay times $\tau_d \lesssim 1$ $\mu$s would generate measurable currents $I \gtrsim 1$ fA through the double dot system. Delay times of such a magnitude are within reach with radio frequency detection techniques.

The gray, shaded regions indicate again where the second law of thermodynamics is seemingly violated. In these regions, the system acts as Maxwell's demon. Because noise induces heating, the point where $\langle \! \langle Q \rangle \! \rangle = 0$ is pushed towards smaller $\Gamma \tau_d$ when $\sigma$ is increased. For larger $\sigma$, there is thus a smaller region where the system is operated as a Maxwell demon. For larger $\Gamma$, power is still produced. However, the energy source is no longer provided by heat but by the voltage gates $W_g$.  In this regime, the system therefore acts as an electron pump \cite{pump1,Martinis-PRL-1994,Keller-APL-1996,pump2}.

%As $\Gamma$ is increased, $\langle \! \langle Q \rangle \! \rangle$ and $\langle \! \langle W_g \rangle \! \rangle$ are growing rapidly in opposite directions. This happens as mistakes in the operation occur more frequently for larger $\Gamma$. E.g.~if an electron tunnels in and out from the left dot in level setting $\{\epsilon_0,\epsilon_u\}$, the levels can mistakenly be changed to $\{\epsilon_l,\epsilon_l\}$. An electron entering the left dot in the latter setting might tunnel to the right dot and reside there. If the energy level setting then is changed to $\{\epsilon_u,\epsilon_0\}$, the electron can enter the right reservoir. During this process, the gates would apply work on the system, i.e.~$\langle \! \langle W_g \rangle \! \rangle < 0$, and if the electron leaves the left dot after the energy level transit, the right reservoir would be heated, thereby $\langle \! \langle Q \rangle \! \rangle >0$. Hence, for large tunneling rates, the system operation starts to resemble an electron pump \cite{pump1,Martinis-PRL-1994,Keller-APL-1996,pump2}.

\section{Conclusions \& outlook}

We presented an implementation of Maxwell's demon in a double quantum dot system. Under ideal operation conditions, i.e. infinitely fast measurement and feedback, electrons are transported against a voltage bias without performing any net work on the electrons. Thus, information is used to convert heat into electrical work. The distributions of produced work and heat are found to be non-Gaussian.

By means of a Monte Carlo method, we have simulated the system under non-ideal operation conditions, i.e.~having delay and noise in the occupation measurements of the dots. Delay slows the operation down, decreasing the produced electrical power, while noise mainly results in heating. The electron trajectories can be fully resolved when having small electron tunneling rates, and the system can be operated in the ideal regime. Slightly increasing the rates, the system is still operated as a Maxwell demon, but it is no longer expected to recover the ideal power production. Further increasing the tunneling rates, the electron trajectories can no longer be resolved, and the operation starts to resemble an electron pump. 

Promising avenues to pursue include the quantum regime, where backaction of the measurement has to be taken into account, as well as the quantum-to-classical transition.

\begin{acknowledgments}
This work was supported by the Swedish Research Council. P.P.P.
acknowledges funding from the European Union's Horizon 2020 research and
innovation programme under the Marie Sk{\l}odowska-Curie Grant Agreement
No. 796700.
\end{acknowledgments}

\appendix

\section{Monte Carlo simulation}
\label{sec:appendix:details-monte-carlo}

\subsection{Rate equation}

The system state $S(t)$ is governed by the rate equation $\boldsymbol{\dot{\rho}}(t) = \mathcal{L} \left[ D(t) \right] \boldsymbol{\rho}(t)$, where the Liouvillian $\mathcal{L} \left[ D(t) \right]$ is dependent on the level setting $D(t) = 0,1,2$,
\begin{equation}
\mathcal{L} \left[ 0 \right] = 
\begin{pmatrix}
- \gamma^{(i)}_{L}(\epsilon_0) \! - \! \gamma^{(i)}_{R}(\epsilon_u) & \gamma^{(o)}_{L}(\epsilon_0) & \gamma^{(o)}_{R}(\epsilon_u) \\
\gamma^{(i)}_{L}(\epsilon_0) & -\gamma^{(o)}_{L}(\epsilon_0) & 0 \\
\gamma^{(i)}_{R}(\epsilon_u) & 0 & - \gamma^{(o)}_{R}(\epsilon_u)
\end{pmatrix}
,
\label{eq:appendix-L0}
\end{equation}
\begin{equation}
\begin{aligned}
& \mathcal{L} \left[ 1 \right] = \\
& \begin{pmatrix}
- \gamma^{(i)}_{L}(\epsilon_l) \! - \! \gamma^{(i)}_{R}(\epsilon_l) & \gamma^{(o)}_{L}(\epsilon_l) & \gamma^{(o)}_{R}(\epsilon_l) \\
\gamma^{(i)}_{L}(\epsilon_l) & - \gamma^{(o)}_{L}(\epsilon_l) \! - \!\gamma & \gamma \\
\gamma^{(i)}_{R}(\epsilon_l) & \gamma & - \gamma^{(o)}_{R}(\epsilon_l) \! - \! \gamma
\end{pmatrix}
,
\end{aligned}
\label{eq:appendix-L1}
\end{equation}
\noindent and
\begin{equation}
\mathcal{L} \left[ 2 \right] = 
\begin{pmatrix}
- \gamma^{(i)}_{L}(\epsilon_u) \! - \! \gamma^{(i)}_{R}(\epsilon_0) & \gamma^{(o)}_{L}(\epsilon_u) & \gamma^{(o)}_{R}(\epsilon_0) \\
\gamma^{(i)}_{L}(\epsilon_u) & -\gamma^{(o)}_{L}(\epsilon_u) & 0 \\
\gamma^{(i)}_{R}(\epsilon_0) & 0 & - \gamma^{(o)}_{R}(\epsilon_0)
\end{pmatrix}
,
\label{eq:appendix-L2}
\end{equation}
\noindent where the transition rates are given by
\begin{equation}
\gamma^{(i)}_{\alpha}(\epsilon_j) \! = \! \Gamma_{\alpha} f_{\alpha}\left( \epsilon_j \right), \hspace{5mm} \gamma^{(o)}_{\alpha}(\epsilon_j) \! \! = \! \! \Gamma_{\alpha} \left[ 1 \! - \! f_{\alpha}\left( \epsilon_j \right) \right].
\label{eq:appendix:rates}
\end{equation} 
\noindent The superscripts $(i)$ and $(o)$ denote whether a transition is into or out from the double dot, the subscript $\alpha$ specifies whether reservoir $L$ or $R$ is involved, and $j=l,0,u$ defines at which energy the transition occurs.

\subsection{Simulation}
To simulate the system state at each point in time, we exploit that under a small change of time $\delta t$, where $D(t)$ is constant, it is possible to write 
\begin{equation}
\boldsymbol{\rho}(t+\delta t) = \left( \mathbb{1} + \mathcal{L}\left[ D(t)\right] \delta t \right) \boldsymbol{\rho}(t).
\label{eq:appendix:sim-eq}
\end{equation}
\noindent The change in time $\delta t$ should be small enough such that 
\begin{equation}
\gamma_{\alpha}^{(i/o)} (\epsilon_j) \delta t \ll 1.
\end{equation}
Given $S(t)=j$ and $D(t)=k$, with $j,k=0,1,2$, at time $t$, we assign $S(t+\delta t)=i$, $i=0,1,2$, with probability 
\begin{equation}
P\left[ S(t+\delta t)=i | S(t)\! = \! j, D(t) \! = \! k \right] = \delta_{ij} + \delta t \left( \mathcal{L}[k] \right)_{ij},
\end{equation}
\noindent which follows from the probability vector $\boldsymbol{\rho}(t+\delta t)$ in Eq.~(\ref{eq:appendix:sim-eq}). Here, $\delta_{ij}$ is the Kronecker delta, and $\left( \mathcal{L}[k] \right)_{ij}$ is the matrix elements of the Liouvillian given in Eqs.~(\ref{eq:appendix-L0})-(\ref{eq:appendix-L2}).
Before the next time step of the simulation, $d(t+\delta t)$ and $D(t+\delta t)$ must be calculated: First, $m(t+\delta t)$ is computed by Eq.~(\ref{eq:mean-of-d(t)}). Then a random number is drawn according to Eq.~(\ref{eq:distribution-detector-output}) and assigned to $d(t+\delta t)$. Finally, $D(t+\delta t)$ is determined by Eq.~(\ref{eq:feedback-for-setting}).

\bibliography{draft1}

\end{document}